\documentclass[a4paper,aps,onecolumn,nofootinbib]{revtex4}
\usepackage[english]{babel}
\usepackage[latin1]{inputenc}
\usepackage[T1]{fontenc}
\usepackage{amsmath}
\usepackage{amsfonts}
\usepackage{amssymb}
\usepackage{graphicx}
\usepackage{multirow} 
\def\rf#1{(\ref{#1})}

\def\de#1/de#2{\frac{\partial {#1}}{\partial {#2}}}

\newcommand{\A}{\mathbb{A}}

\newcommand{\ba}{\begin{eqnarray}}
\newcommand{\ea}{\end{eqnarray}}
\newcommand{\be}{\begin{equation}}
\newcommand{\ee}{\end{equation}}

\def\DerN#1{\frac{d #1}{d N}}
\def\G{{\cal G}}
\DeclareMathOperator{\sech}{sech}
\begin{document}
%%%%%%%%%%%%%%%%%%%%%%%%%%%%%%%%%%%%%%%%%%%%%%%%%%%%%%%%%%%%%%%%%%%%%%%%%%%%%%%%%%%%%%%%%%%%%%%%%%%
\title{Phase space of modified Gauss-Bonnet gravity.}
\author{Sante Carloni \footnote{E-mail: sante.carloni@tecnico.ulisboa.pt}}
\affiliation{Centro Multidisciplinar de Astrof\'{\i}sica - CENTRA,
Instituto Superior Tecnico - IST,
Universidade de Lisboa - UL,
Avenida Rovisco Pais 1, 1049-001, Portugal}
\author{Jos\'{e} P. Mimoso \footnote{E-mail: jpmimoso@fc.ul.pt}}
\address{Faculdade de Ci\^{e}ncias, Departamento de F\'{\i}sica
\& Instituto de Astrof\'{\i}sica e Ci\^encias do Espa\c co, Universidade de Lisboa,
Ed. C8, Campo Grande, 1749-016 Lisboa,
Portugal}
\date{\today}
%%%%%%%%%%%%%%%%%%%%%%%%%%%%%%%%%%%%%%%%%%%%%%%%%%%%%%%%%%%%%%%%%%%%%%%%%%%%%%%%%%%%%%%%%%%%%%%%%%%
\begin{abstract}
We investigate the evolution of  non vacuum Friedmann-Lema\^{\i}tre-Robertson-Walker (FLRW) with any spatial curvature in the context of Gauss-Bonnet gravity. The analysis employs a new method which enables us to  explore the phase space of any specific theory of this class. We consider several examples, discussing the transition from a decelerating into an acceleration universe within these theories. We also deduce from the dynamical equations some general conditions on the form of the action which guarantee the presence of specific behaviours like the the emergence of accelerated expansion. As in $f(R)$ gravity, our analysis shows that there is a set of initial conditions for which these models have a finite time singularity which can be an attractor. The presence of this instability  also in  the Gauss-Bonnet gravity is to be ascribed to the fourth-order derivative in the field equations, i.e., is the direct consequence of the higher order of the equations. 

\end{abstract}
%%%%%%%%%%%%%%%%%%%%%%%%%%%%%%%%%%%%%%%%%%%%%%%%%%%%%%%%%%%%%%%%%%%%%%%%%%%%%%%%%%%%%%%%%%%%%%%%%%%
\maketitle

%%%%%%%%%%%%%%%%%%%%%%%%%%%%%%%%%%%%%%%%%%%%%%%%%
\section{Introduction}
%%%%%%%%%%%%%%%%%%%%%%%%%%%%%%%%%%%%%%%%%%%%%%%%
The search for a purely geometrical description  of dark energy has 
 led the research community to the analysis of a number of 
possible extensions of General Relativity (GR). From the now ``classical'' $f(R)$ \cite{f(R) review} and scalar tensor \cite{ScTNBooks} theories to more complex extensions which involve the presence of torsion or of more complicated invariants in the gravitational actions \cite{Clifton:2011jh}. These theories have been thoroughly studied and have revealed interesting features as well as a number of problems connected to different kinds of instabilities. Among those theories which include terms of { fourth order} in the derivative of the metric, the so-called modified Gauss-Bonnet gravity has been shown to have interesting properties. 

The idea of these theories comes from  the concept of brane words, which are in turn derived from string theories. In five dimensions one considers a Gauss-Bonnet (GB) term which is normally non minimally coupled with a scalar (modulus or dilaton) field. One can shown that the induced theory on four dimensions is able  to generate the de Sitter solution %as well as have 
and has  other relevant cosmological properties \cite{Calcagni:2006ye,Deruelle}. 

In \cite{Od} a new class of theories of gravity, dubbed Gauss-Bonnet gravity, was proposed in which the Gauss-Bonnet term appears in four dimensions. As it is well known, the Gauss-Bonnet invariant is a total divergence in four dimensions and therefore a linear GB term would be irrelevant in this case. The new class of theories  overcomes this issue  introducing in the Hilbert-Einstein action a generic non linear function of the Gauss-Bonnet term. The cosmology of Gauss-Bonnet gravity has been thoroughly studied \cite{Cognola:2006eg,Nojiri:2007te,Bamba:2014mya,Garcia:2010xz}. Some of the members of this class of theories were shown to be able to pass the Solar System tests %as well as 
and to possess cosmological solutions %able to represent 
 producing cosmic acceleration \cite{DeFelice:2008wz,DeFelice:2009aj,Davis:2007id}. The same authors also found that linear cosmological pertubations %on 
 of a spatially flat background  contains some instabilities \cite{DeFelice:2009rw}. Although this is clearly a problem, one must remember that % higher order theories dynamics 
the dynamics of higher order theories
depends on the value of the spatial curvature in a more complex way that standard (GR). This means that even if $k$ is almost zero, as recently found by  the Planck collaboration \cite{Ade:2015xua}, the dynamical features of these models can be completely different, especially at perturbative level (see eqs. \rf{CosmEqVar} below for a glimpse of this).  This fact justifies a further analysis of these theories which takes %in 
into account spatial curvature.

Gauss-Bonnet gravity, like many other extensions of GR, has a structure that makes it difficult to understand the physics of these models. In this respect, the use of semi quantitative techniques, like the dynamical systems approach (DSA) \cite{DSA}, can help in %clarify 
{ clarifying} details of the dynamics of cosmological models based on these theories  and other modifications of GR \cite{ModGravDSA} which are not obvious. The use of DSA for this purpose is now fairly standard, however the analysis is often limited to basic aspects of the method, like for example the choice of variables and the closure do the dynamical system. Recently, a new strategy to alleviate these  problems was proposed in the context of $f(R)$-gravity \cite{Carloni:2015jla}.  With respect to the previously proposed approaches, the new method has the advantage to be applicable to any form of the function $f$ and leads to a clearer (albeit not complete)  understanding of expanding cosmologies. 

In this work, we will propose a similar technique to treat Gauss Bonnet cosmologies in which the action is the sum of the Hilbert- Einstein  term and a generic (non linear) function of the Gauss Bonnet invariant. As in the case of $f(R)$ gravity the method will allow to treat any forms of the function $f$ and will enable us to characterise the nature of the attractors (if any) for the cosmologies Gauss-Bonnet gravity. We will apply this technique to a number of versions of this theory that have been deemed interesting and/or compatible  with the present data. The results of { this} study will be the starting point for a more complete analysis of the dynamics of linear perturbations, which will be pursued elsewhere.

Unless otherwise specified, natural units ($\hbar=c=k_{B}=G=1$)
will be used throughout this paper, Latin indices run from 0 to 3.
The symbol $\nabla$ represents the usual covariant derivative and
$\partial$ corresponds to partial differentiation. We use the
$(+,-,-,-)$ signature and the Riemann tensor is defined by
\begin{equation}
R^{a}{}_{bcd}=W^a{}_{bd,c}-W^a{}_{bc,d}+ W^e{}_{bd}W^a{}_{ce}-
W^f{}_{bc}W^a{}_{df}\;,
\end{equation}
where the $W^a{}_{bd}$ are the Christoffel symbols (i.e. symmetric in
the lower indices), defined by
\begin{equation}
W^a_{bd}=\frac{1}{2}g^{ae}
\left(g_{be,d}+g_{ed,b}-g_{bd,e}\right)\;.
\end{equation}
The Ricci tensor is obtained by contracting the {\em first} and the
{\em third} indices
\begin{equation}\label{Ricci}
R_{ab}=g^{cd}R_{acbd}\;.
\end{equation}
Finally the Hilbert--Einstein action in the presence of matter is
given by
\begin{equation}
S=\int d x^{4} \sqrt{-g}\left[\frac{1}{2\kappa}R+ L_{m}\right]\;,
\end{equation}
where $\kappa =8\pi$ and has the dimension of the inverse of a length square.

%%%%%%%%%%%%%%%%%%%%%%%%%%%%%%%%%%%%%%%%%%%%%%%%%
\section{Basic Equations}
%%%%%%%%%%%%%%%%%%%%%%%%%%%%%%%%%%%%%%%%%%%%%%%%
The Action for modified Gauss-Bonnet gravity reads 
\begin{equation}
S=\int d^4x \sqrt{-g}\left[\frac{R}{2\kappa}+f({\cal G})\right]+S_M(g^{\mu\nu},\psi)\,,
   \label{modGBaction}
\end{equation}
where $S_M(g^{\mu\nu},\psi)$ is the matter
action and $\psi$ collectively denotes the matter
fields. The Gauss-Bonnet invariant is defined as
\begin{equation}
{\cal G}\equiv
R^2-4R_{\mu\nu}R^{\mu\nu}+R_{\mu\nu\alpha\beta}R^{\mu\nu\alpha\beta}.
\end{equation}

Varying the action with respect to the metric yields the field equations:
\begin{eqnarray}
\label{geeq}
G_{\mu \nu}+8 \left[ R_{\mu \rho \nu \sigma} +R_{\rho \nu} g_{\sigma \mu}
-R_{\rho \sigma} g_{\nu \mu} -R_{\mu \nu} g_{\sigma \rho}+
R_{\mu \sigma} g_{\nu \rho}+\tfrac{R}{2} (g_{\mu \nu} g_{\sigma \rho}
-g_{\mu \sigma} g_{\nu \rho}) \right] \nabla^{\rho} \nabla^{\sigma} f'
+(\G f'-f) g_{\mu \nu}=\kappa T_{\mu \nu}\,,
\end{eqnarray}
where the prime represnts the derivative with respect to $\G$ and $G_{\mu \nu}=R_{\mu \nu}-(1/2)R g_{\mu \nu}$ is the Einstein-tensor.  

The matter energy-momentum tensor is defined as usual as
\begin{equation}
T_{\mu \nu }=-\frac{2}{\sqrt{-g}}\frac{\delta (\sqrt{-g}\,{L}_{m})}{\delta
(g^{\mu \nu })} \,.  \label{EMTdef2}
\end{equation}

Our treatment will consider  Friedmann Lema\^{\i}tre Robertson Walker (FLRW) metric:
\begin{equation}\label{frw}
 ds^2 = -dt^2 + a^2(t)\left[ \frac{dr^2}{1-kr^2} + r^2 (d\theta^2 +
\sin^2\theta d\phi^2)\right]\;,
\end{equation}
where $a$ is the scale factor and  $k$ the spatial curvature. We also assume that the cosmic fluid is a prefect fluid 
with equation of state $p=w \mu$  with $0\leq w\leq1$, where $\mu$ and  $p$, respectively are the energy density and pressure measured by the co-moving observer, and where we assume   $w$ to be constant. In this metric, the Gauss-Bonnet invariant is 
\begin{equation}
\G=24\left(H^2+\frac{k}{a^2}\right)(\dot{H}+H^2)
\end{equation}
where $H=\frac{\dot{a}}{a}$ is the Hubble factor. This expression is particularly important because it connects the sign of $\G$ to the sign of the deceleration factor. For $k\geq0$, any cosmology which transits between acceleration and deceleration will have to change the sign of $\G$. If $k<0$, this is not necessarily true. In this respect Gauss-Bonnet cosmology depends in an even more crucial way on the spatial curvature than GR.

%%%%%%%%%%%%%%%%%%%%%%%%%%%%%%%%%%%%%%%%%%%%%%%%%
\section{The Dynamical Systems Approach.}
%%%%%%%%%%%%%%%%%%%%%%%%%%%%%%%%%%%%%%%%%%%%%%%
First, we introduce a  constant $\chi_0$ such that the products $R\chi_0$ and ${\cal G}\chi_0^2$ are dimensionless and we  redefine all the constants appearing in the action \rf{modGBaction} to obtain
\begin{equation}\label{action}
S=\int \left\{\chi_0 R+ f\left({\cal G} \chi_0^2\right)+{\ L}%
_{m}\left(g_{\mu \nu},\Phi\right)\right\} \sqrt{-g}\;d^{4}x~
\end{equation}%
where now all the constants (with the exception of $\chi_0$) in the action are assumed dimensionless.

Now defining the parameters
\begin{equation} \label{HubbleVarDS}
{\mathfrak q} =\frac{\dot{H}}{H^2},\quad {\mathfrak j} =\frac{\ddot{H}}{ H^3}-\frac{\dot{H}^2}{
   H^4},\quad{\mathfrak s}=\frac{\dddot{H}}{ H^4}+3\frac{\dot{H}^3}{ H^6}-4\frac{ \dot{H}
   \ddot{H}}{ H^5}\,.
\end{equation}
the cosmological equations can be written as
\begin{align}\label{CosmEqVar}
&\nonumber 3\chi_0\left(H^2+ \frac{k}{a^2}\right) - \mu +f -24 H^2 (\mathfrak{q}+1) \left( H^2+\frac{k}{a^2}\right) f'+\left\{3^2\,2^6  H^4 \left(\mathfrak{j}+\mathfrak{q}^2-2\right)\frac{k^2}{a^4} \right.\\
&~~~~~~~~~~~~~~~~~~~~~\left.+3^2\,2^7 H^6  \left(5 \mathfrak{j}+2 \mathfrak{q} (3+1)
-9\right)\frac{k}{a^2}+2^6 3^4 H^8 [\mathfrak{j}+\mathfrak{q} (3 \mathfrak{q}+4)]\right\} f''=0\\
&\nonumber \chi_0 H^2 (\mathfrak{q}+1)+\frac{1}{3} \mu  (3 w+1)+\frac{f}{3}-24 H^2 (\mathfrak{q}+1) \left(\frac{k}{a^2}+H^2\right) f'\\
& \nonumber +\left\{2^5 3^2  H^4  \left(\frac{k}{a^2}+H^2\right)
   \left(\frac{k}{a^2}+9 H^2\right)\mathfrak{s}+2^5 3^2  H^8 [\mathfrak{j} (92 \mathfrak{q}+29)+ \mathfrak{q}^2 (87 \mathfrak{q}+131)-28\mathfrak{q}]\right.\\
&\nonumber \left. +2^5 3^2 H^4 [\mathfrak{j} (4 \mathfrak{q}-3)+(\mathfrak{q}-3) \mathfrak{q}^2+6(\mathfrak{q}-1) ] \frac{k^2}{a^4}+2^6 3^2 H^6  [\mathfrak{j} (24
   \mathfrak{q}-11)+\mathfrak{q} (2 \mathfrak{q} (5 \mathfrak{q}-3)-31)+25]\frac{k}{a^2}\right\}f''\\
 &\nonumber  + \left\{2^8 3^3 H^6  \left(\mathfrak{j}+\mathfrak{q}^2-2\right)^2\frac{k^3}{a^6}+2^8 3^3 H^8 
   \left(\mathfrak{j}+\mathfrak{q}^2-2\right) [11 \mathfrak{j}+\mathfrak{q} (15 \mathfrak{q}+8)-18]\frac{k^2}{a^4}\right.\\
 &\left.  +2^8 3^3 H^{10} [\mathfrak{j}+\mathfrak{q} (3 \mathfrak{q}+4)] [19
   \mathfrak{j}+\mathfrak{q} (21 \mathfrak{q}+4)-36] \frac{k}{a^2}+2^8 3^5  H^{12} [\mathfrak{j}+\mathfrak{q} (3 \mathfrak{q}+4)]^2\right\}f'''=0
\end{align}
Here the non trivial role of the spatial curvature is particularly evident: the
$k=0$ equations are very different from the full ones.
 
In order to analyse the phase space let us define the expansion normalized variables 
\begin{align} \label{DynVar}
\begin{split}
&\mathbb{G}=\frac{\G}{3 H^4},\quad \mathbb{K}=\frac{k}{a^2 H^2},\quad \Omega =\frac{\mu }{3H^2 \chi_0},\\
& \mathbb{J}=\mathfrak j\quad \mathbb{Q}=\mathfrak q,\quad \mathbb{A}=\chi_0H^2\,.
\end{split}
\end{align}
We also  define the logarithmic (dimensionless) ``time variable'' $N=\ln a$. Note that in choosing this time variable we are assuming that we represent the phase space for $H>0$, i.e., we are considering only expanding cosmologies.

It is important to stress that, since $\mathbb{G}$ has the same sign of $\G$, a sufficient condition for the transition between deceleration and deceleration can only be obtained if $\mathbb{G}=0$ is not an invariant submanifold of the phase space.

The phase space associated to the equations \rf{CosmEqVar} is then described by the autonomous system
\begin{equation}\label{DynSys}
\begin{split}
&\DerN{\mathbb{G}}= \frac{\mathbb{G}}{2}  \left(8-\frac{\mathbb{G}}{\mathbb{K}+1}\right)-\frac{8
   [\mathbb{K}+\mathbf{X}-\mathbb{G} \mathbf{Y}-\Omega +1]}{(\mathbb{K}+9) \mathbf{Z}},\\ 
&\DerN{\Omega} =\Omega  \left[\frac{\mathbb{G}}{4 (\mathbb{K}+1)}+(3 w+1)\right],\\ 
&\DerN{\mathbb{K} }=-\frac{\mathbb{K}\, \mathbb{G} }{4
   \mathbb{K}+4}\\
&\DerN{\mathbb{A}}=\frac{\mathbb{A}}{4}  \left(\frac{\mathbb{G}}{\mathbb{K}+1}-8\right)\,,
\end{split}
\end{equation}
together with the two constraints
\begin{eqnarray}
&& 0=(9+\mathbb{K}) \left[\mathbb{J}+\mathbb{K}
   \left(\mathbb{J}+\mathbb{Q}^2-2\right)\mathbf{Z}+\mathbb{Q} (3
   \mathbb{Q}+4)\right]+\mathbb{K}+\mathbf{X}-\mathbb{G} \mathbf{Y}-\Omega +1  \label{FriedConstr} \\
&&  \mathbb{G}=8(1+\mathbb{\mathbb{K}})(1+\mathbb{Q}).\label{GB Constr}
\end{eqnarray}
 In all the equations above we have defined
\begin{align} \label{XYT}
& {\bf X}= \frac{f}{3 \chi_0 H^2 }\\
&  {\bf Y}= \frac{f' H^2}{\chi_0}\\
& {\bf Z}= \frac{3\, 2^6 H^6  f''}{ \chi_0}
\end{align}
Note that choosing the variables above, the system is singular for $\mathbb{K}+1=0$. Such singularity is originated by our choice of coordinates:  selecting for example the variable $\mathbb{Q}$ instead of $\mathbb{G}$ it can be eliminated to obtain the system
\begin{equation}\label{DynSysQ}
\begin{split}
&\DerN{\mathbb{Q}}=-\mathbb{Q}^2-\frac{ (\mathbb{Q}^2-2) \mathbb{K}+ \mathbb{Q}(3\mathbb{Q}+4)}{(\mathbb{K}+1)}-\frac{\mathbb{K}-\Omega+1+{\bf X}-8 (\mathbb{K}+1) (\mathbb{Q}+1) {\bf Y} }{(\mathbb{K}+1) (\mathbb{K}+9) {\bf Z}},\\ 
&\DerN{\Omega} =-\Omega  \left[2 \mathbb{Q}+3( w+1)\right],\\ 
&\DerN{\mathbb{K} }=-2\mathbb{K}\, (1+\mathbb{Q})\\
&\DerN{\mathbb{A}}=\mathbb{A}\mathbb{Q}\,,
\end{split}
\end{equation}
where now ${\bf X},{\bf Y},{\bf Z}$ are function of $\mathbb{Q},\mathbb{K}$. The two systems are equivalent when one is away from $\mathbb{K}=-1$. Since in the examples we consider there is no special point for $\mathbb{K}=-1$, and moreover there is no appreciable difference in the structure of the fixed points,  we will perform the analysis using Eqs.~\rf{DynSys}. However one must stress that $\mathbb{K}=-1$ 
is nevertheless of interest because, since both $\mathbb{G}$ and its derivatives become zero, 
this case represents a state in which the theory  effectively becomes of order two.  Therefore the system \rf{DynSysQ} can be useful to explore this property of Gauss-Bonnet gravity.
 
The solutions associated to the fixed points can be derived using the modified Raychaudhury equation (the second of the  Eqs.~\rf{CosmEqVar}) in the variables \rf{HubbleVarDS}. In a fixed point we have
\begin{equation}\label{mathfraks}
\begin{split}
&\frac{1}{H}\frac{ d^3{H}}{d N^3}=\mathfrak{s}_*=-\frac{1}{(K_*+1) (K_*+9) {\bf Z}_*} \left\{2 (K_*+9) {\bf T}_* \left[\mathbb{J}_*+K_* \left(\mathbb{J}_*+\mathbb{Q}_*^2-2\right)+\mathbb{Q}_* (3 \mathbb{Q}_*+4)\right]^2+2 \mathbb{Q}_* \right.\\
&\left.~~~~~+(1+3 w) \Omega_*+2 {\bf X}_*-2 \mathbb{G}_* {\bf Y}_*+2+
  29{\bf Z}_* \mathbb{J}_*+2{\bf Z}_* \mathbb{Q}_* [46 \mathbb{J}_*+K_* (2 \mathbb{J}_* (K_*+12)-3 K_*-31)]\right.\\
&\left.~~~~~+K_*{\bf Z}_* [50-22 \mathbb{J}_*-3 \mathbb{J}_* K_*+6 K_*]+{\bf Z}_* [K_* (K_*+20)+87]
   \mathbb{Q}_*^3+{\bf Z}_*[131-3 K_* (K_*+4)] \mathbb{Q}_*^2-28{\bf Z}_* \mathbb{Q}_*\right\}
 \end{split}
\end{equation}
where
\begin{equation}
{\bf T}= \frac{2^8 3^2H^{10}  f'''}{ \chi_0}
\end{equation}
and an asterisk represents the value of a quantity at a fixed point.  The general solution for the above equation is
\begin{equation}\label{SolHGen} 
H=\left\{ 
\begin{array}{ll}
H_0+H_1 N+ H_2 N^2 & \mathfrak{s}^*=0\\
H_0 e^{-p N}+ e^{\frac{p N}{2}}\left(H_1\cos \frac{pN\sqrt{3}}{2}+ H_2 \sin \frac{pN\sqrt{3}}{2}\right)&\mathfrak{s}^*\neq 0
\end{array}
\right.
\end{equation}
where $p=-\sqrt[3]{\mathfrak{s}^*}$ and the $H_i$ are integration constants. Using the definition of $N$ the above equations translate in equations for the scale factor
\begin{equation}
\frac{\dot{a}}{a}=\left\{ 
\begin{array}{ll}
H_0+H_1 \ln a+ H_2 (\ln a)^2 & \mathfrak{s}^*=0\\
H_0 a^{-p}+ a^{p/2}\left[H_1\cos \left( \frac{p\sqrt{3}}{2}\ln a\right)+ H_2 \sin \left( \frac{p\sqrt{3}}{2}\ln a\right)\right]&\mathfrak{s}^*\neq 0
\end{array}
\right.
\end{equation}
In the fist case the equation can be solved exactly to give
\begin{equation}\label{SolH}
  a(t)=a_0 \exp \left\{\frac{\sqrt{4 H_2 H_0-H_1^2}}{2 H_2} \tan \left[\frac{1}{2} (t-t_0) \sqrt{4 H_2
   H_0-H_1^2}\right]-\frac{H_1}{2 H_2}\right\}.
\end{equation}
which was already found in the case of $f(R)$-gravity \cite{Carloni:2015jla}. 

A major problem one finds when these solutions are found is 
associated with the determination of the integration constants. Such constants are important as they  
concur to determine the nature of the solution. For example, the solution given by \rf{SolH} will have a finite time singularity if $4 H_2 H_0-H_1^2>0$. One way in which the value of these constants can be inferred is to consider the value of the initial condition of the orbits that approach this point.
%%%%%%%%%%%%%%%%%%%%%%%%%%%%%%%%%%%%%%%%%%%%%%%%%%
\section{Examples}
%%%%%%%%%%%%%%%%%%%%%%%%%%%%%%%%%%%%%%%%%%%%%%%%%
We will now apply the machinery presented above to a number of physically relevant theories. We will then compare our results with 
 those found in literature.
%%%%%%%%%%%%%%%%%%%%%%%%%%%%%%%%%%%%%%%%%%%%%%%%%%
\subsection{${\cal G}^n$-gravity}
%%%%%%%%%%%%%%%%%%%%%%%%%%%%%%%%%%%%%%%%%%%%%%%%%
As a first example, let us consider the model 
\begin{equation}
f= \alpha \chi_0^{2n} {\cal G}^n 
\end{equation}
 we have 
 \begin{align} \label{XYT}
& {\bf X}= 3^{n-1} \alpha  \mathbb{A}^{2 n-1} \mathbb{G}^n ,\\
&  {\bf Y}=3^{n-1} n \alpha  \mathbb{A}^{2 n-1} \mathbb{G}^{n-1}\\
& {\bf Z}=2^6\ 3^{n-1} (n-1) n \alpha  \mathbb{A}^{2n-1} \mathbb{G}^{n-2}\\
&  {\bf T}= 2^8\ 3^{n-1} (n-2) (n-1) n \alpha  \mathbb{A}^{2 n-1} \mathbb{G}^{n-3}\,.
\end{align}
where the constraint  \rf{GB Constr} holds. The dynamical equations read
\begin{equation}\label{DynSysEx1}
\begin{split}
&\DerN{\mathbb{G}}= \mathbb{G} \left[4-\mathbb{G}\left(\frac{1 }{\mathbb{K}+1}+\frac{1}{4n(\mathbb{K}+9) } \right)+\frac{2^{2-n}
   3^{1-n} \mathbb{A}^{1-2 n} (\mathbb{K}-\Omega +1) \mathbb{G}^{1-n}}{n(1-n) \alpha (\mathbb{K}+9)
   }\right],\\ 
&\DerN{\Omega} =\Omega  \left[\frac{\mathbb{G}}{4 (\mathbb{K}+1)}+(3 w+1)\right],\\ 
&\DerN{\mathbb{K} }=-\frac{\mathbb{K}\, \mathbb{G} }{4(
   \mathbb{K}+1)}\\
&\DerN{\mathbb{A}}=\frac{\mathbb{A}}{4}  \left(\frac{\mathbb{G}}{\mathbb{K}+1}-8\right)\,,
\end{split}
\end{equation}
The system admits four invariant submanifolds ($\mathbb{K}=0, \Omega=0,\mathbb{A}=0, \mathbb{G}=0$). These results imply that no global attractor exists in general.

The fixed subspaces for this system are indicated in Table \ref{TableFPEx1}. We have a line of fixed points $\mathcal{L}$ for $\mathbb{K}=\mathbb{K}_0, \Omega=0,\mathbb{A}=0, \mathbb{G}=0$  for $2+3n<0$. Other fixed points exist only 
at specific intervals of $n$. For example, the point  $\mathcal{A}$  is present for $2-3n>0$, whereas $\mathcal{B}$ is present only for  $(n-1)\alpha>0$.

\begin{figure}[htbp]
\begin{center}
\includegraphics[scale=0.7]{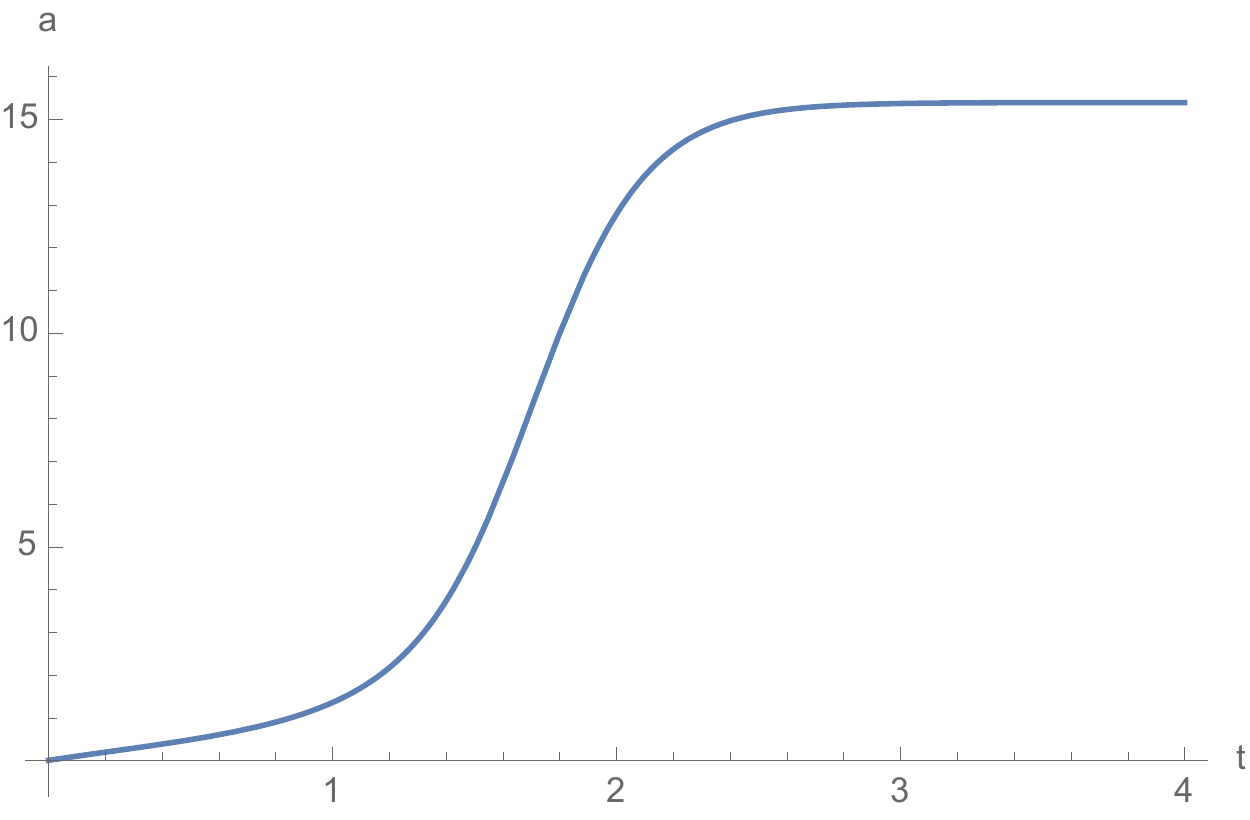}
\caption{Numerical solution of equation \rf{SolA}. The constants $H_i$ have all been chosen to be one and the initial condition is $a(0)=0.01$.}
\label{PlotSolA}
\end{center}
\end{figure}

The points on the line $\mathcal{L}$ are associated 
with the solution given by the equation
\begin{equation}\label{SolA}
\frac{\dot{a}}{a}=\frac{H_1}{a}+a^{1/2} \left[H_2 \sin \left(\frac{1}{2}\sqrt{3} \log a\right)+H_3 \cos\left(\frac{1}{2}\sqrt{3} \log a\right)\right].
\end{equation}
For point $\mathcal{A}$  we have 
\begin{equation}\label{SolB}
\frac{\dot{a}}{a}=H_1a^\frac{1-36 n}{\sqrt[3]{2^9 3^3 n^2+1}}+a^{1/2} \left[H_2 \sin \left(\frac{(1-36 n)\sqrt{3}}{\sqrt[3]{2^{10} 3^3 n^2+1}}  \log a\right)+H_3 \cos\left(\frac{(1-36 n)\sqrt{3}}{\sqrt[3]{2^{10} 3^3 n^2+1}} \log a\right)\right].
\end{equation}
whereas  $\mathcal{B}$  corresponds to solution \rf{SolH}. A numerical integration of the equation \rf{SolA} and \rf{SolB} is given in Figures \ref{PlotSolA} and \ref{PlotSolB}.

The stability of all the fixed points  can be deduced using the standard Hartmann-Grobmann theorem~\cite{Hartman-Grobman} and it is indicated in Table \ref{TableFPEx1}. The points on $\mathcal{L}$ are all unstable and $\mathcal{A}$ and $\mathcal{B}$ can be attractors or saddles depending on the values of $n$ and $\alpha$. 

Because of the presence of the invariant submanifold there is no orbit that can represent a transition between decelerated and accelerated cosmologies. In this respect this theory is not useful to model dark energy and could only be used as patch model for eternal inflation. Even in this case, however, these models can have the same issues of $f(R)$ gravity, i.e. the onset of a finite time singularity.  Our conclusions are consistent with the results in \cite{DeFelice:2009aj,Davis:2007id}. 

%%%%%%%%%%%%%%%%%%%%%%%%%%%%%%%%%%%%%%%%%%%%%%%%%%%%%%%%
%%%%%%%%%%%%%%%%%%%%%%%%%%%%%%%%%%%%%%%%%%%%%%%%%%%%%%%%
%%%%%%%%%%%%%%%%%%%%%%%%%%%%%%%%%%%%%%%%%%%%%%%%
%%%%%%%%
\begin{figure}[htbp]
\begin{center}
\includegraphics[scale=0.7]{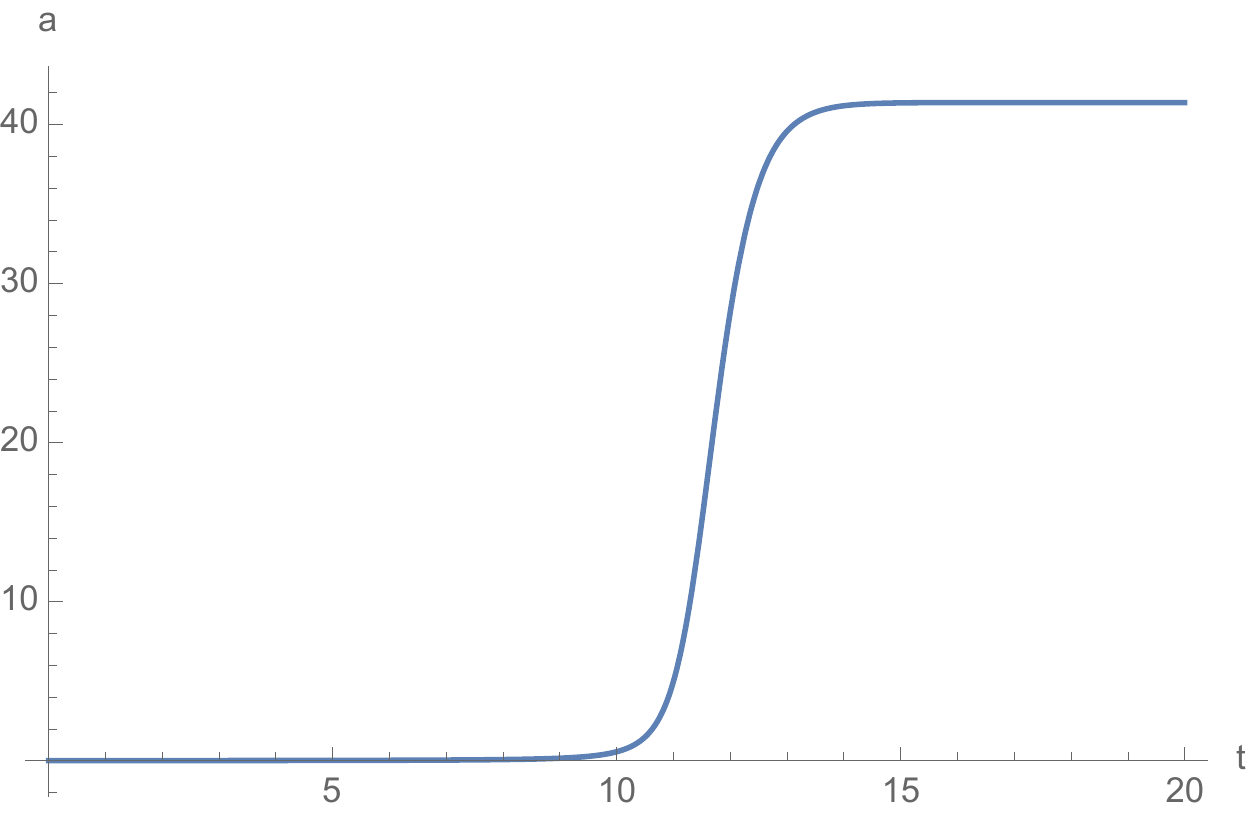}
\caption{Numerical solution of equation \rf{SolB}. The constants $H_i$ have all been chosen to be one and the initial condition is $a(0)=0.01$.}
\label{PlotSolB}
\end{center}
\end{figure}
%%%%%%%%%%%%%%%%%%%%%%%%%%%%%%%%%%%%%%%%%%%%%%%%%%%%%%%%
%%%%%%%%%%%%%%%%%%%%%%%%%%%%%%%%%%%%%%%%%%%%%%%%
%%%%%%%%
\begin{figure}[htbp]
\begin{center}
\includegraphics[scale=0.4]{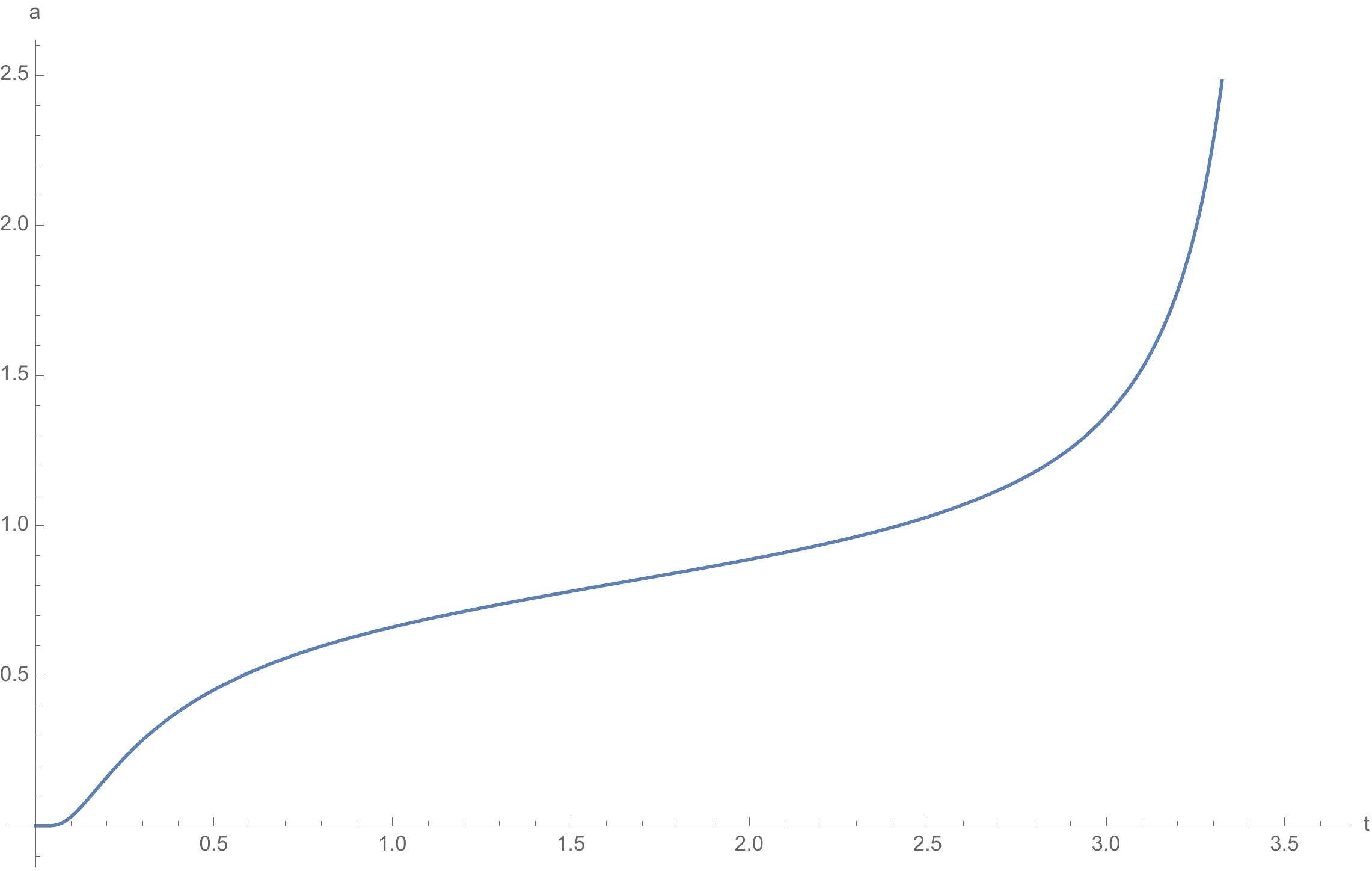}
\caption{Numerical solution of equation \rf{SolH}. The constants $H_i$ have all been chosen in such a way that  $4 H_0 H_2-H_1^2>0$ and the initial condition is $a(0)=0.01$. The solution presents a finite time singularity. If the above condition is violated the solution approaches a static universe as in the case of \rf{SolA} an \rf{SolB}. }
\label{PlotSolB}
\end{center}
\end{figure}
%%%%%%%%%%%%%%%%%%%%%%%%%%%%%%%%%%%%%%%%%%%%%%%%%%%%%%%%

%%%%%%%%%%%%%%%%%%%%%%%%%%%%%%%%%%%%%%%%%%%%%%%%%%%%%%%%
\begin{table}
\caption{Fixed subspaces of $f({\cal G})=\alpha {\cal G}^{n}$ and their  associated solutions. Here $a_0=H_1+H_2+H_3$, $\mathbb{A}_*$ is the solution of the algebraic equation A stays for attractor, F$_A$ for attractive focus, S for saddle. } \label{TableFPEx1}
\begin{tabular}{llccc} \hline\hline
~~~~~~~~~~~& Coordinates $\{\mathbb{G},\mathbb{K},\Omega, \A\}$  & Scale Factor&Existence  & Stability\\ \hline\\
$\mathcal{L}$ & $\left\{0, \mathbb{K}_0, 0 , 0\right\}$ & Solution of \rf{SolA} &$2-3n>0$ & Unstable  \\ \\
\multirow{2}{*}{$\mathcal{A}$} & \multirow{2}{*}{$\left\{  \frac{2^5 3^2}{36 n-1},0, 0 ,0\right\}$ } & \multirow{2}{*}{Solution of \rf{SolB}} & \multirow{2}{*}{$2-3n>0$}& A $n<0$\\ 
&&&&S $0<n<2/3$ \\ \\
\multirow{3}{*}{$\mathcal{B}$} & \multirow{3}{*}{$\left\{8, 0, 0 , \left[3^{n-1} 8^n (n-1) \alpha \right]^{\frac{1}{1-2 n}}\right\}$}  & \multirow{3}{*}{\rf{SolH}} & \multirow{3}{*}{$(n-1) \alpha >0$} & A $\frac{72}{1369}\leq n<\frac{1}{2}$ \\
&&&& F$_A$  $0<n<\frac{72}{1369}$\\ 
&&&& S  otherwise\\ \\ 
\hline\hline\\
 \end{tabular}
\end{table}
 %%%%%%%%%%%%%%%%%%%%%%%%%%%%%%%%%%%%%%%%%%%%%%%%%%%%%%%%%%%

%%%%%%%%%%%%%%%%%%%%%%%%%%%%%%%%%%%%%%%%%%%%%%%%%%
\subsection{The De Felice-Tsujikawa models.}
%%%%%%%%%%%%%%%%%%%%%%%%%%%%%%%%%%%%%%%%%%%%%%%%%
In Ref.~\cite{DeFelice:2008wz} some particular forms of $f(G)$ 
were proposed which,  based on exact and perturbative arguments, 
 were considered to  be cosmologically viable. In our formulation such models can be written as
\begin{equation}
\begin{split}
& f(\G)= \alpha \chi_0^2 \G \arctan\left(\chi_0^2 \G\right)-\beta \chi_0,\\
& f(\G)= \alpha \chi_0^2 \G \arctan\left(\chi_0^2 \G\right) -\beta \chi_0-\gamma \ln\left(1+\chi_0^2 \G\right),\\
& f(\G)= \alpha \ln\left[\cosh\left(\chi_0^2 \G\right)\right]-\beta \chi_0,
\end{split}
\end{equation}
where $\beta$, differently form $\alpha$ and $\gamma$, is a dimensional constant. All these constant are considered positive. The presence of $\beta$ requires the introduction of a new variable $\mathbb{B}= \beta/ H^2$ whose dynamic equation is
\begin{equation}
\DerN{\mathbb{B}}=-\frac{3 \mathbb{B} (\mathbb{G}-8 \mathbb{K}-8)}{4
   (\mathbb{K}+1)}.
\end{equation}
By including this additional equation in the system \rf{DynSys}, we are ready to explore these models.

%%%%%%%%%%%%%%%%%%%%%%%%%%%%%%%%%%%%%%%%%%%%%%%%%%
\subsubsection{Moldel 1: $f(\G)= \alpha \chi_0^2 \G \arctan\left(\chi_0^2 \G\right)-\beta \chi_0$}
%%%%%%%%%%%%%%%%%%%%%%%%%%%%%%%%%%%%%%%%%%%%%%%%%
For the first form of $f$ above we have 
 \begin{align} \label{XYT}
& {\bf X}=\alpha  \mathbb{A} \mathbb{G} \arctan\left(3 \mathbb{A}^2
   \mathbb{G}\right)- \mathbb{B},\\
&  {\bf Y}=\alpha  \mathbb{A}
   \left[\arctan\left(3 \mathbb{A}^2 \mathbb{G}\right)+\frac{3 \mathbb{A}^2
   \mathbb{G}}{9 \mathbb{A}^4 \mathbb{G}^2+1}\right]\\
& {\bf Z}=\frac{\alpha\, 3\,  2^7 \, 
   \mathbb{A}^3}{\left(9 \mathbb{A}^4 \mathbb{G}^2+1\right)^2}\\
&  {\bf T}= -\frac{\alpha 2^{11}\, 3^3   \mathbb{A}^7 \mathbb{G}}{\left(9 \mathbb{A}^4
   \mathbb{G}^2+1\right)^3}\,.
\end{align}
and the dynamical equations read
\begin{equation}\label{DynSysEx2a}
\begin{split}
&\DerN{\mathbb{G}}= \frac{1}{2} \mathbb{G}
   \left(8-\frac{\mathbb{G}}{\mathbb{K}+1}\right)+\frac{9 \mathbb{A}^4 \mathbb{G}^2+1}{48 \alpha  \mathbb{A}^3 (K+9)}\left\{1+K-\Omega -\mathbb{B}-3 \mathbb{A}^3 \mathbb{G}^2 \left[\alpha +3 \mathbb{A} (\mathbb{B}+\Omega )\right]+9
   \mathbb{A}^4 \mathbb{G}^2 (K+1)\right\}, \\
&\DerN{\Omega} =\Omega  \left[\frac{\mathbb{G}}{4 (\mathbb{K}+1)}+(3 w+1)\right],\\ 
&\DerN{\mathbb{K} }=-\frac{\mathbb{K}\, \mathbb{G} }{4(
   \mathbb{K}+1)},\\
&\DerN{\mathbb{A}}=\frac{\mathbb{A}}{4}  \left(\frac{\mathbb{G}}{\mathbb{K}+1}-8\right)\,,\\
&\DerN{\mathbb{B}}=-\frac{3 \mathbb{B} (\mathbb{G}-8 \mathbb{K}-8)}{4
   (\mathbb{K}+1)}.
\end{split}
\end{equation}
The system admits three invariant submanifolds ($\mathbb{K}=0, \Omega=0,\mathbb{A}=0$). The last one, depending on the values of the parameters, can be singular. These results imply that no global attractor exists in general, but they allow the possibility of a transition between decelerated and accelerated cosmologies.

The system \rf{DynSysEx2a}  presents only a two dimensional fixed point subspace with coordinates
\begin{equation}
\mathcal{S}=\{\mathbb{G}_*,\mathbb{K}_*,\Omega_*, \A_*, \mathbb{B}_*\}=\left\{8,0,0, \A_*, 1-\frac{192 \alpha  \mathbb{A}_*^3}{576 \mathbb{A}_*^4+1}\right\}
\end{equation}
which corresponds to the solution \rf{SolH}. Note that, by definition $\mathbb{B}>0$ and this implies that this subspace can exist only for $\alpha> \frac{576 A_*^4+1}{192 A_*^3}$.

\begin{figure}[htbp]
\begin{center}
\includegraphics[scale=0.7]{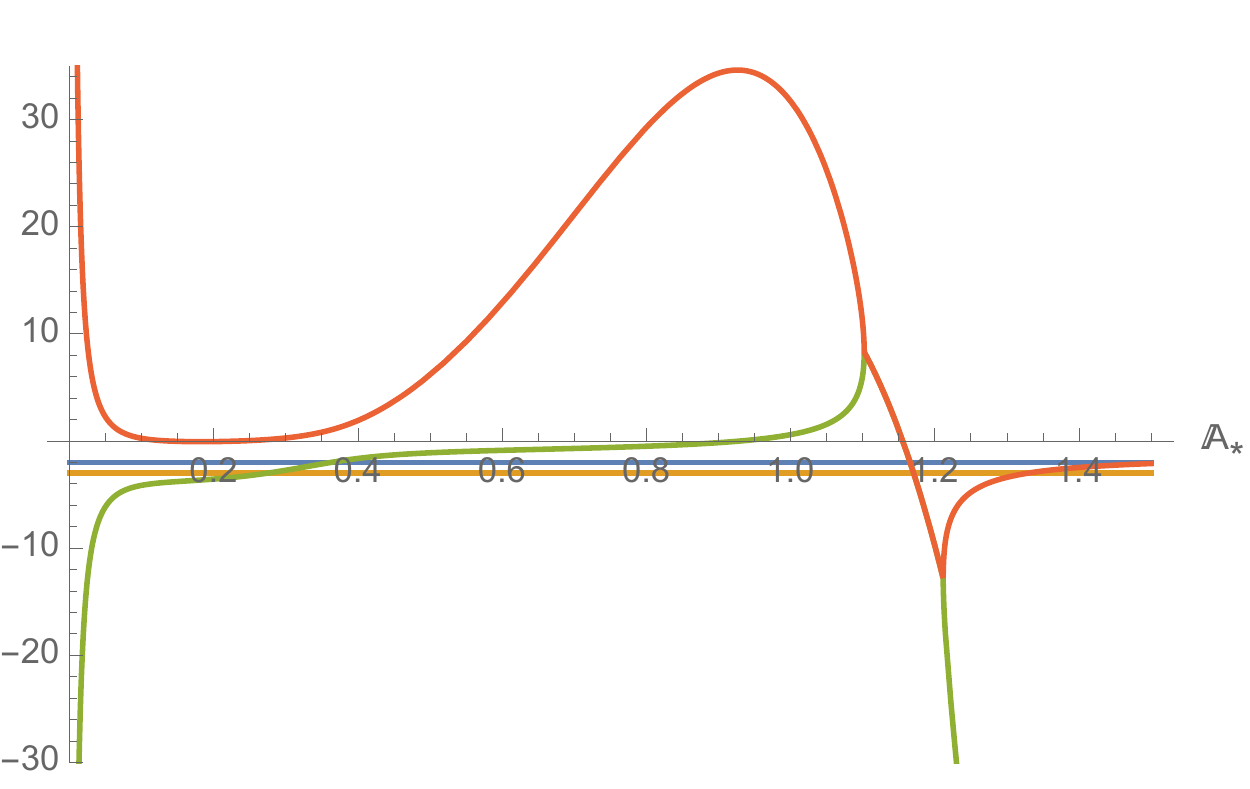}
\caption{Plot of the real part of the eigenvalues of the fixed points of the line $\mathcal{L}$ for $f(\G)= \alpha \chi_0^2 \G \arctan\left(\chi_0^2 \G\right)-\beta \chi_0^2$. Here $\alpha$ has been chosen to be $3$ and $w=0$. }
\label{EigenEx1DFT}
\end{center}
\end{figure}

The stability of the fixed points on the surface  $\mathcal{S}$ can be obtained using the Hartmann-Grobmann theorem~\cite{Hartman-Grobman}.  A plot of the real part of the eigenvalues is given in Figure \ref{EigenEx1DFT}. It is evident that the point on  $\mathcal{S}$ can be saddles or attractors depending on the values of $\alpha$.  An easier way to visualize the phase space dynamics is to define the variable 
\begin{equation}
\mathbb{Y}=\mathbb{B}-\frac{3 A \left(576 \mathbb{A}^4-192 \alpha  \mathbb{A}^3+1\right)}{\mathbb{A}^2 \left(576 \mathbb{A}^4+1\right)}
\end{equation}
and analyse the corresponding dynamical system:
\begin{equation}\label{DynSysEx2aY}
\begin{split}
&\DerN{\mathbb{G}}=  \frac{1}{2} \mathbb{G}
   \left(8-\frac{\mathbb{G}}{\mathbb{K}+1}\right)+\frac{9 \mathbb{A}^4 \mathbb{G}^2+1}{48 \alpha  \mathbb{A}^3 (K+9)}\left\{1+K-\Omega -3 \mathbb{A}^3 \mathbb{G}^2 \left(\alpha +3 \mathbb{A} \Omega \right)+9
   \mathbb{A}^4 \mathbb{G}^2 (K+1)\right\} \\
   &~~~~~~~~~+\frac{\left(9 A^4 \mathbb{G}^2+1\right)^2 }{ A (K+9)}\left(\frac{ A \mathbb{Y}+3}{48 \alpha  A^3}-\frac{12 }{576 A^4+1}\right),\\
&\DerN{\Omega} =\Omega  \left[\frac{\mathbb{G}}{4 (\mathbb{K}+1)}+(3 w+1)\right],\\ 
&\DerN{\mathbb{K} }=-\frac{\mathbb{K}\, \mathbb{G} }{4(
   \mathbb{K}+1)},\\
&\DerN{\mathbb{A}}=\frac{\mathbb{A}}{4}  \left(\frac{\mathbb{G}}{\mathbb{K}+1}-8\right)\,,\\
&\DerN{\mathbb{Y}}=\frac{3 (-\mathbb{G}+8 K+8)}{4 \mathbb{A} \left(576 \mathbb{A}^4+1\right)^2 (K+1)} \left[\left(576 \mathbb{A}^4+1\right)^2 (\mathbb{A} \mathbb{Y}+2)-192 \alpha  \mathbb{A}^3 \left(576
   \mathbb{A}^4+5\right)\right] .
\end{split}
\end{equation}
which presents only %a
one  line of fixed points ($\left\{8,0,0, \A_*, 0\right\}$) rather than a surface and shows clearly that there is no heteroclinic orbit that connects the unstable points to the stable points on the line. This implies that there cannot be an orbit in which  an unstable phase characterised by \rf{SolH} is followed by a stable phase characterized by the same solution.

This implies that the theory has an accelerated expansion attractor, but in order to conclude that a transition deceleration/acceleration is possible one should look at the numerical evolution of the orbits.  This result is consistent with what was found in \cite{DeFelice:2009aj}, in which one of such transitions has been observed by numerical integration.

There is no general way to deduce the size of the set of initial conditions that leads to cosmic acceleration, however, by inspection, we can deduce that the smaller the value of the parameter $\alpha$ the larger the initial conditions that lead to the cosmic acceleration transition.

In this model the presence of the  cosmological constant does not seem to change too much the dynamics. When $\mathbb{B}=0$ the fixed subspace reduces to a one dimensional space of fixed points associated to the solution \rf{SolH}. Their character can be once again  either  that of a saddle or  of an attractor, depending on the value of $\alpha$. Again the bigger the parameter $\alpha$ the lower the number of stable point in the one dimensional space. The only concrete difference between this model and the more general one is the fact that the equations to solve are easier in the presence of the cosmological constant.
%%%%%%%%%%%%%%%%%%%%%%%%%%%%%%%%%%%%%%%%%%%%%%%%
%%%%%%%%
%%%%%%%%%%%%%%%%%%%%%%%%%%%%%%%%%%%%%%%%%%%%%%%%%%%%%%%%
 %%%%%%%%%%%%%%%%%%%%%%%%%%%%%%%%%%%%%%%%%%%%%%%%%%
\subsubsection{Model 2: $f(\G)= \alpha \chi_0^2 \G \arctan\left(\chi_0^2 \G\right)-\gamma \ln\left(1+\chi_0^2 \G\right)  -\beta \chi_0$}
%%%%%%%%%%%%%%%%%%%%%%%%%%%%%%%%%%%%%%%%%%%%%%%%%
For this form of $f$ above we have 
 \begin{align} \label{XYT}
& {\bf X}=-\frac{\gamma  \log \left(9 \mathbb{A}^4
   \mathbb{G}^2+1\right)}{3 \mathbb{A}}+\alpha  \mathbb{A} \mathbb{G} \tan ^{-1}\left(3
   \mathbb{A}^2 \mathbb{G}\right)- \mathbb{B},\\
&  {\bf Y}=\alpha 
   \mathbb{A} \tan ^{-1}\left(3 \mathbb{A}^2 \mathbb{G}\right)+\frac{3 \mathbb{A}^3
   \mathbb{G} (\alpha -2 \gamma )}{9 \mathbb{A}^4 \mathbb{G}^2+1}\\
& {\bf Z}=\frac{384 \mathbb{A}^3 \left(\alpha +\gamma  \left(9 \mathbb{A}^4
   \mathbb{G}^2-1\right)\right)}{\left(9 \mathbb{A}^4
   \mathbb{G}^2+1\right)^2}\\
&  {\bf T}= -\frac{27648 \mathbb{A}^7 \mathbb{G}
   \left(2 \alpha +9 \mathbb{A}^4 \gamma  \mathbb{G}^2-3 \gamma \right)}{\left(9
   \mathbb{A}^4 \mathbb{G}^2+1\right)^3}\,.
\end{align}
and the dynamical equations read
\begin{equation}\label{DynSysEx2b}
\begin{split}
&\DerN{\mathbb{G}}= \frac{1}{2} \mathbb{G}
   \left(8-\frac{\mathbb{G}}{\mathbb{K}+1}\right)+\frac{9 \mathbb{A}^4 \mathbb{G}^2+1}{48 \alpha  \mathbb{A}^3 (K+9)}\left\{1+K-\Omega -\mathbb{B}-3 \mathbb{A}^3 \mathbb{G}^2 \left[\alpha +3 \mathbb{A} (\mathbb{B}+\Omega )\right]+9
   \mathbb{A}^4 \mathbb{G}^2 (K+1)\right\}\\
   &~~~~~~~~~+\frac{\gamma  \left(9 A^4 \mathbb{G}^2+1\right)^2 }{144 A^4 (K+9) \left(\alpha-\gamma +9 A^4 \gamma  \mathbb{G}^2
   \right)}\ln \left(9 A^4
   \mathbb{G}^2+1\right),\\ 
&\DerN{\Omega} =\Omega  \left[\frac{\mathbb{G}}{4 (\mathbb{K}+1)}+(3 w+1)\right],\\ 
&\DerN{\mathbb{K} }=-\frac{\mathbb{K}\, \mathbb{G} }{4(
   \mathbb{K}+1)},\\
&\DerN{\mathbb{A}}=\frac{\mathbb{A}}{4}  \left(\frac{\mathbb{G}}{\mathbb{K}+1}-8\right)\,,\\
&\DerN{\mathbb{B}}=-\frac{3 \mathbb{B} (\mathbb{G}-8 \mathbb{K}-8)}{4
   (\mathbb{K}+1)}.
\end{split}
\end{equation}
The system above differs for the previous one only by just one term. It has essentially the same general features in terms of invariant submanifold and fixed points.  The two dimensional fixed point subspace has coordinates
\begin{equation}
\mathcal{L}=\{\mathbb{G}_*,\mathbb{K}_*,\Omega_*, \A_*, \mathbb{B}_*\}=\left\{8,0,0, \A_*, 1-\frac{192 \alpha  \mathbb{A}_*^3}{576 \mathbb{A}_*^4+1}+\gamma  \left[\frac{384
   \mathbb{A}_*^3}{576 \mathbb{A}_*^4+1}-\frac{\ln \left(576 \mathbb{A}_*^4+1\right)}{3
   \mathbb{A}_*}\right]\right\}
\end{equation}
the solution associate to these points is \rf{SolH}.

Using the Hartmann-Grobmann theorem it is easy to prove that, also in this case, the fixed points on the line can be either attractors or saddles, with the difference that now the stability depends on $\gamma$ other than $\alpha$.  Therefore the same picture of the previous model seems to emerge in this case. The logarithmic correction seems to be irrelevant on Friedmannian dynamics, but will surely be relevant at  perturbative level.

%%%%%%%%%%%%%%%%%%%%%%%%%%%%%%%%%%%%%%%%%%%%%%%%%%
\subsubsection{Model 3: $f(\G)=\alpha \ln\left[\cosh\left(\chi_0^2 \G\right)\right]-\beta \chi_0$}
%%%%%%%%%%%%%%%%%%%%%%%%%%%%%%%%%%%%%%%%%%%%%%%%%
For this form of $f$ above we have 
 \begin{align} \label{XYT}
& {\bf X}=\frac{\alpha} {3 \mathbb{A}} \ln \left[\cosh \left(3 \mathbb{A}^2 \mathbb{G}\right)\right]-
   \mathbb{B},\\
&  {\bf Y}=\alpha  \mathbb{A} \tanh \left(3 \mathbb{A}^2 \mathbb{G}\right) ,\\
& {\bf Z}=3\alpha\, 2^6 \mathbb{A}^3
   \text{sech}^2\left(3 \mathbb{A}^2 \mathbb{G}\right),\\
&  {\bf T}= -2^9 3^2 \alpha  \mathbb{A}^5 \tanh \left(3 \mathbb{A}^2 \mathbb{G}\right)
   \sech^2\left(3 \mathbb{A}^2 \mathbb{G}\right)\,.
\end{align}
and the dynamical equations read
\begin{equation}\label{DynSysEx2b}
\begin{split}
&\DerN{\mathbb{G}}= \frac{1}{2}
   \mathbb{G} \left(8-\frac{\mathbb{G}}{\mathbb{K}+1}\right)+\frac{\cosh ^2\left(3 \mathbb{A}^2 \mathbb{G}\right) \left\{3 \mathbb{A}( \mathbb{B}-K+ \Omega -1)-\alpha  \ln\left[\cosh \left(3 \mathbb{A}^2 \mathbb{G}\right)\right]\right\}}{72
   \alpha  \mathbb{A}^4 (K+9)}\\ 
& ~~~~~~~~~+\frac{\mathbb{G} \sinh \left(3 \mathbb{A}^2 \mathbb{G}\right) \cosh \left(3 \mathbb{A}^2
   \mathbb{G}\right)}{24 \mathbb{A}^2 (K+9)}\\
&\DerN{\Omega} =\Omega  \left[\frac{\mathbb{G}}{4 (\mathbb{K}+1)}+(3 w+1)\right],\\ 
&\DerN{\mathbb{K} }=-\frac{\mathbb{K}\, \mathbb{G} }{4(
   \mathbb{K}+1)},\\
&\DerN{\mathbb{A}}=\frac{\mathbb{A}}{4}  \left(\frac{\mathbb{G}}{\mathbb{K}+1}-8\right)\,,\\
&\DerN{\mathbb{B}}=-\frac{3 \mathbb{B} }{2} \left(\frac{\mathbb{G}}{\mathbb{K}+1}-8\right).
\end{split}
\end{equation}
The system admits  again three invariant submanifolds: $\mathbb{K}=0, \Omega=0,\mathbb{A}=0$, and the latter can be singular. These results imply that, in general, no global attractor exists also for this model.

The system \rf{DynSysEx2b} presents only a two dimensional fixed point subspace with coordinates
\begin{equation}
\mathcal{L}=\{\mathbb{G}_*,\mathbb{K}_*,\Omega_*, \A_*, \mathbb{B}_*\}=\left\{8,0,0, \A_0, 1-8 \alpha\mathbb{A}_* \tanh \left(24 \mathbb{A}_*^2\right)+\frac{\alpha  \log \left(\cosh \left(24
   \mathbb{A}_*^2\right)\right)}{3 \mathbb{A}_*}\right\}
\end{equation}
as in the case of Model 2, the solution associated to this fixed point is  \rf{SolH}.

\begin{figure}[htbp]
\begin{center}
\includegraphics[scale=0.7]{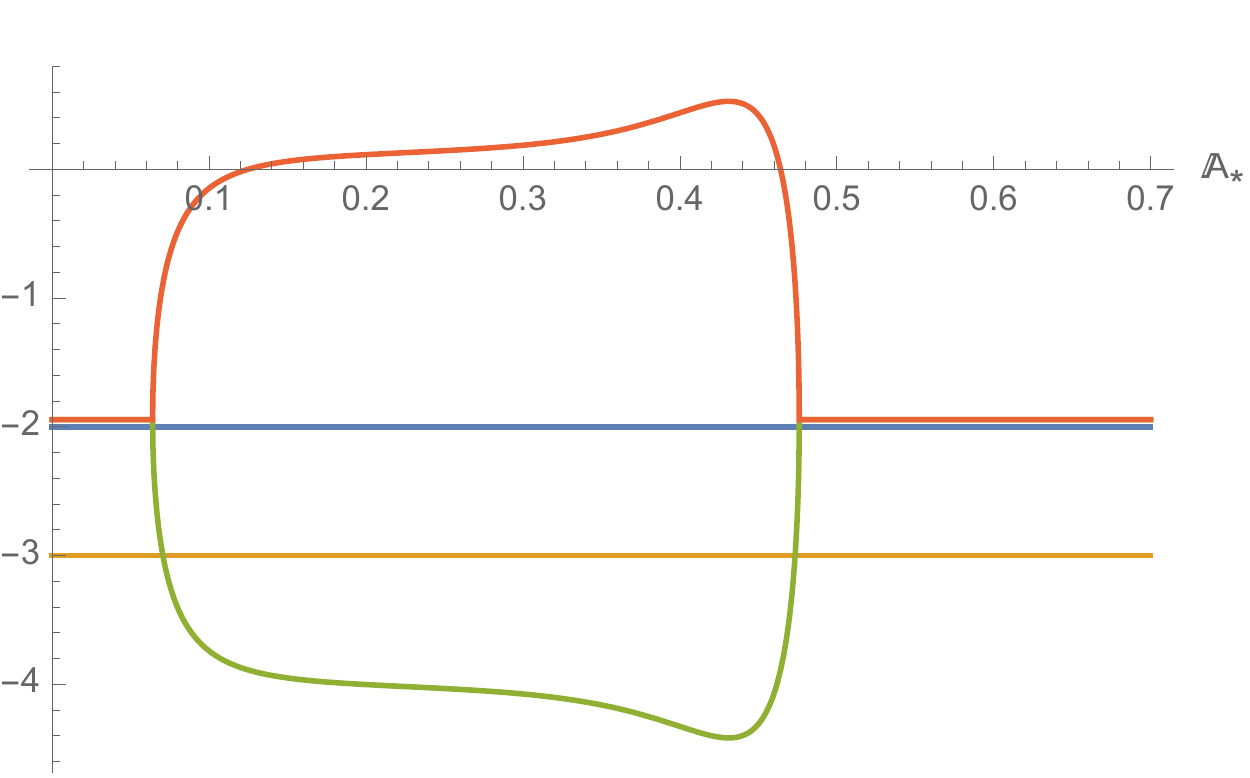}
\caption{Plot of the real part of the eigenvalues of the fixed points of the line $\mathcal{L}$ for Model 3. Here $\alpha$ has been chosen to be $3$ and $w=0$. Its increase wides the set of values of $\mathbb{A}_*$ for which two of the eigenvalues have discordant sign.}
\label{EigenEx3DFT}
\end{center}
\end{figure}

The stability of the fixed points is of the same type to the two previous models, depending essentially on the value of the variable $\alpha$. In Figure \ref{EigenEx3DFT} we  
 represent a plot of the eigenvalues of these points.

In spite of the different analytical  form, the  observed dynamics essentially is  the same as  in the previous two models. This should not be surprising as all these models have been explicitly designed to have the same characteristics. Our results confirm this fact. As for  model 2, differences between the behaviour of these cosmologies will likely become evident at perturbative level.
%%%%%%%%%%%%%%%%%%%%%%%%%%%%%%%%%%%%%%%%%%%%%%%%
%%%%%%%%
%%%%%%%%%%%%%%%%%%%%%%%%%%%%%%%%%%%%%%%%%%%%%%%%%%%%%%%%

\section{Conclusions}
%%%%%%%%%%%%%%%%%%%%%%%%%%%%%%%%%%%%%%%%%%%%%%%
In this paper we have proposed a new formalism to analyse the cosmology of Gauss-Bonnet  gravity. The method allows us to explore the phase space of any theory, and associates to the fixed points solutions for the scale factor  which present four integration constants. 

Many features of this class of theories become clear  by looking at the structure of dynamical equations. For example, a transition between decelerating and  accelerating cosmology can only happen if the submanifold $\mathbb{G}=0$ is not an invariant submanifold, and therefore the dynamical equations can be used as election tools for this class of theories, even without the actual phase space analysis. More specifically,  we can conclude that all theories for which   $\mathcal{G}f'(\mathcal{G})-f(\mathcal{G})\neq1$ ( i.e. $f$ is non linear) and $f''(\mathcal{G})^{-1}=0$ for $\mathcal{G}\rightarrow0$ will have  an invariant submanifold. In other words, the transition between deceleration and acceleration requires a nonlinear function whose second derivative does not diverge in $\mathcal{G}$. This is a general validity criteria of any model of Gauss-Bonnet  gravity, and it is fulfilled by the models of section IVB. In fact this condition overlaps with those already found in \cite{DeFelice:2008wz}. Our analysis shows also that there are conditions for which another type of fixed point can appear, which could play the role of %decelerated 
decelerating fixed point. However, from the first of  Eqs.~(\ref{DynSys}) we can deduce that the condition for the existence of this point is precisely the same as the one  
 for the existence of the invariant submanifold. This implies that even if this point could be a  fixed point  associated with a decelerated expansion there would not be heteroclinic orbits
connecting it to an accelerated expansion one. 

We have applied the new technique to four different models. One of them $f(\mathcal{G})=\alpha \mathcal{G}^n$ has been considered for its simplicity and as a testing ground for the new method.  The presence of the invariant manifold $\mathbb{G}=0$ forbids the presence of a transition within acceleration and deceleration if $n<2$.  Depending on the sign of $n$, there could be two types of attractors one characterised by the solution \rf{SolA} ($n<0$) and the other by \rf{SolB} ($n>0$). Since this last solution can lead to a big rip scenario, one can conclude that $n<0$ is a more physically interesting scenario for this model. It is interesting to note that for these values of $n$ the constraints from measurement in the Solar System are known to be much less strict \cite{Davis:2007id}.

Next we studied three examples of the models proposed in \cite{DeFelice:2008wz}. They all have very similar phase spaces. This is to be expected from the way in which these models are constructed, and thus gives a confirmation of the consistency of the method we have used. In accordance with the results available in literature, in this case we find that these models admit a transition
from decelerated to accelerated expansion, but that this transition and the approach to the accelerated expansion attractor depend on the initial conditions, and are  
not guaranteed.

Finally, as in $f(R)$ gravity  and because of the higher order of the equations, our analysis shows that there is a set of initial conditions for which these models have a finite time singularity which is an attractor. The presence of this instability also in the Gauss-Bonnet gravity under consideration  implies that the existence  of this instability is to be ascribed to the fourth-order derivative in the field equations. 

%%%%%%%%%%%%%%%%%%%%%%%%%
\section*{Acknowledgements}
This work was supported by  the Funda\c{c}\~{a}o para a Ci\^{e}ncia e Tecnologia through project IF/00250/2013.  SC also acknowledge financial support provided under the European Union's H2020 ERC Consolidator Grant ``Matter and strong-field gravity: New frontiers in Einstein's theory'' grant agreement no.  MaGRaTh-646597, and under the H2020-MSCA-RISE-2015 Grant No. StronGrHEP-690904. JPM  acknowledges the financial  support by Funda\c{c}\~ao para a Ci\^encia e a Tecnologia (FCT) through the research grant UID/FIS/04434/2013. The authors also acknowledge the COST Action
CA15117, supported by COST (European Cooperation in Science and
Technology). 

%%%%%%%%%%%%%%%%%%%

\end{document}